\documentclass[preprint,12pt]{elsarticle}

\usepackage{amssymb}
\usepackage{amsmath}
\usepackage{multirow}
\usepackage{multicol}
\usepackage{booktabs}
\usepackage{graphicx}
\usepackage{subcaption}

\journal{Computers in Biology and Medicine}

\begin{document}

\begin{frontmatter}



\title{A Progressive Risk Formulation for Enhanced Deep Learning based Total Knee Replacement Prediction in Knee Osteoarthritis}


\author[cds_nyu]{Haresh Rengaraj Rajamohan} 
\author[nyulh]{Richard Kijowski} 
\author[cds_nyu]{Kyunghyun Cho} 
\author[nyulh,cbim]{Cem M. Deniz} 

\affiliation[cds_nyu]{organization={Center for Data Science, New York University},
            city={New York},
            postcode={10011}, 
            state={NY},
            country={USA}}
\affiliation[nyulh]{organization={Department of Radiology, New York University Langone Health},
            city={New York},
            postcode={10016}, 
            state={NY},
            country={USA}}
\affiliation[cbim]{organization={Bernard and Irene Schwartz Center for Biomedical Imaging, New York University Langone Health},
            city={New York},
            postcode={10016}, 
            state={NY},
            country={USA}}
\begin{abstract}
We developed deep learning models for predicting Total Knee Replacement (TKR) need within various time horizons in knee osteoarthritis patients, with a novel capability: the models can perform TKR prediction using a single scan, and furthermore when a previous scan is available, they leverage a progressive risk formulation to improve their predictions. Unlike conventional approaches that treat each scan of a patient independently, our method incorporates a constraint based on disease’s progressive nature, ensuring that predicted TKR risk either increases or remains stable over time when multiple scans of a knee are available. This was achieved by enforcing a progressive risk formulation constraint during training with patients who have more than one available scan in the studies. Knee radiographs and MRIs from the Osteoarthritis Initiative (OAI) and Multicenter Osteoarthritis Study (MOST) were used in this work and deep learning models were trained to predict TKR within 1, 2, and 4-year time periods. The proposed approach, utilizing a dual-model risk constraint architecture, demonstrated superior performance compared to baseline - conventional models trained with standard binary cross entropy loss. It achieved an AUROC of 0.87 and AUPRC of 0.47 for 1-year TKR prediction on the OAI radiograph test set, considerably improving over the baseline AUROC of 0.79 and AUPRC of 0.34. For the MOST radiograph test set, the proposed approach achieved an AUROC of 0.77 and AUPRC of 0.25 for 1-year predictions, outperforming the baseline AUROC of 0.71 and AUPRC of 0.19. Similar trends were observed in the MRI testsets.\end{abstract}

\begin{graphicalabstract}
\includegraphics[width=\textwidth]{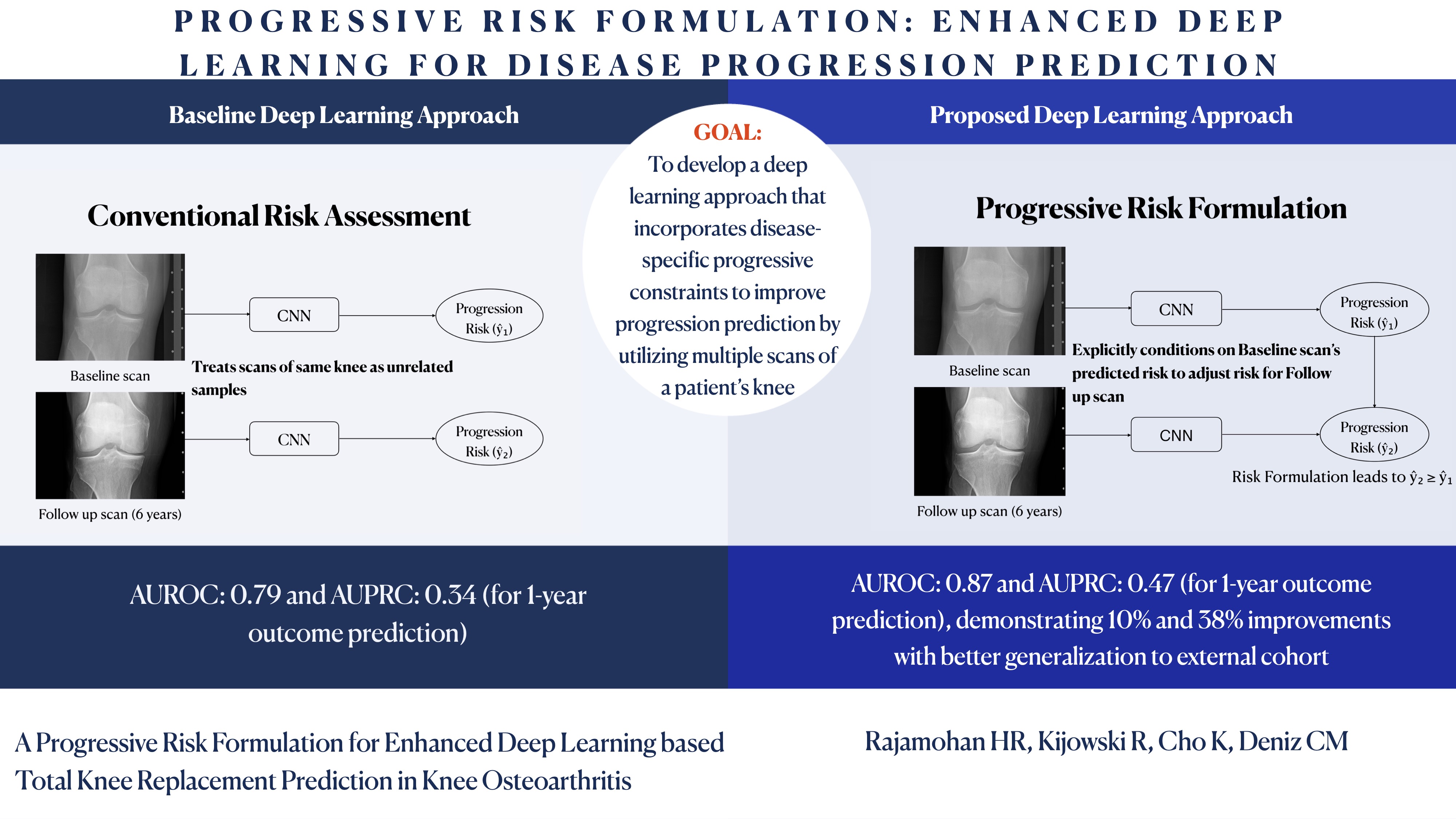}
\end{graphicalabstract}

\begin{highlights}
\item A novel progressive risk formulation that enforces the assumption that Total Knee Replacement (TKR) risk increases or remains stable over time, aligning with the progressive nature of knee osteoarthritis

\item Deep learning models that can make predictions using either a single scan or leverage sequential scans when available by explicitly conditioning on past prediction, improving performance for 1, 2, and 4-year TKR prediction horizons

\item Superior performance metrics with our dual-model architecture (RiskFORM2), achieving AUROC of 0.87 and AUPRC of 0.47 for 1-year TKR prediction, substantially outperforming conventional approaches

\item Enhanced generalizability demonstrated through successful external validation on the MOST dataset, showing robust prediction capabilities across different patient populations

\item Improved predictive performance within each Kellgren-Lawrence grade (KLG), providing clinical value beyond what could be inferred from radiographic grading alone

\end{highlights}

\begin{keyword}
Knee Osteoarthritis \sep OA \sep Total Knee Replacement \sep TKR \sep Kellgren-Lawrence Grade \sep KLG \sep Progression\sep Modified Risk Formulation \sep RiskFORM \sep Deep Learning \sep Convolutional Neural Networks \sep CNN \sep Regularization \sep Machine Learning \sep Risk Prediction \sep Radiography \sep Magnetic Resonance Imaging \sep MRI


\end{keyword}

\end{frontmatter}



\section{INTRODUCTION}
\label{intro}

Osteoarthritis (OA) is the most common form of arthritis and the major cause of physical disability in the elderly. OA causes a major socioeconomic burden with the overall costs associated with the treatment of patients with the disease estimated to be 19,000 Euros per year \cite{cross2014}. OA is generally diagnosed with a combination of clinical symptoms and radiographic findings of osteophyte formation and joint space loss. Predicting OA progression has previously been posed as a supervised learning problem with radiographs of the knee joints as inputs and various OA progression outcomes as labels to be predicted \cite{tolpadi2020,zhang2020,leung2020}. At the endpoint of knee OA progression, total knee joint replacement (TKR) is generally used as the final recourse to relieve pain and restore function in severely diseased knee joints \cite{dougados2009}. Prediction models for TKR outcome have been developed in the past to help determine which knees at baseline have a high risk for undergoing TKR  in the future \cite{tolpadi2020,leung2020,rajamohan2023}. 

Recently, deep learning architectures—particularly convolutional neural networks (CNNs)—have been utilized for knee OA assessment by enabling highly accurate predictions of both disease onset and progression patterns, outperforming traditional statistical models. Tolpadi et al. \cite{tolpadi2020} trained DenseNet \cite{huang2017} models to predict TKR within 5 years, achieving AUROCs of 0.85 for radiographs and 0.89 for MRIs. Tiulpin et al. \cite{tiulpin2018,tiulpin2020} developed CNN models to predict OA based on Kellgren-Lawrence grade (KLG) \cite{kellgren1957}, reaching an impressive AUROC of 0.93. Further, Tiulpin et al. \cite{tiulpin2019a,tiulpin2019b} created multi-modal CNN models to predict knee OA progression within 8 years, achieving an AUROC of 0.75.

For MRI-based approaches, Pedoia et al. \cite{pedoia2019} trained 3D DenseNet models to diagnose knee OA using T2 MRIs achieving AUROC of 0.83, while Rajamohan et al. \cite{rajamohan2023} developed multimodal 3D ResNet models to predict TKR status within 9 years using various MRI contrasts from the OAI datasets and achieved AUROC of 0.89 by ensembling multiple contrasts with clinical information.  Taking a different approach, Guan et al. \cite{guan2022} combined traditional and deep learning approaches to predict OA pain progression, achieving an AUROC of 0.8.

Beyond imaging alone, Ningrum et al. \cite{ningrum2021} leveraged electronic medical records from Taiwan to predict knee OA risk using features such as sequential diagnosis history, drug prescriptions, age, and sex, achieving a high AUROC of 0.95. Kim et al. \cite{kim2020} demonstrated significant performance improvements by incorporating clinical information alongside knee radiographs. Zhang et al. \cite{zhang2020} enhanced KL-grade prediction and generalization using radiographs by implementing attention mechanisms focused on the knee joint. Recent studies have improved these results using various modalities, including radiographs, MRIs, and clinical information for knee OA diagnosis and prognosis [16-30]. Multiple review papers [31-34] have examined the impact of deep learning in this domain, confirming its position as the preferred approach.

Li et al \cite{li2020} used a Siamese architecture with a contrastive loss function to encode the severity of OA using the KL-grade. They found that the Siamese network outputs, which are the Euclidean distances between hidden representations of two knees, correlate well with both longitudinal changes over time in disease severity and across patients with different KL-grades. Their approach, using the pairwise Euclidean distance achieves an AUROC of 0.88 over time and 0.9 across patients. This illustrates the effectiveness of CNN models in encoding OA severity for knee OA diagnosis, which motivated our work.

While CNNs' capability to encode clinically relevant OA features has been demonstrated, existing methods treat longitudinal knee scans as independent samples. In contrast, we propose incorporating domain-specific disease progression patterns through a novel risk-aware prior, enabling the model to learn temporally consistent representations. Knee OA is characterized by a gradual progression of cartilage degeneration and changes in other joint structures. Given its progressive nature, the risk of requiring TKR typically escalates or remains constant over time; a decline in risk is highly unlikely as current treatments are not effective \cite{karsdal2016}. To address this, we propose a novel risk constraint specific to knee OA that enforces the assumption that TKR risk should either increase or remain stable over time. 

During model training, this constraint is applied using pairs of scans from the same knee when available. While the decision to perform TKR involves multiple clinical factors beyond imaging, accurate prediction of future TKR need based on imaging characteristics can aid in early intervention and treatment planning.

Our approach differs from conventional methods in two key aspects: first, it can leverage multiple scans of the same patient during training rather than treating them as independent observations to better understand the disease, and second, it appropriately incorporates the progressive nature of OA through our proposed risk constraint. Importantly, while the risk constraint is applied using paired scans during training, the resulting models can effectively predict TKR risk using a single or sequence of scans during evaluation. Our hypothesis is that this training strategy, which better aligns with the disease's progressive nature, would improve model performance for predicting TKR across different time horizons.

\section{MATERIALS AND METHODS}
\label{mat_methods}

\subsection{Datasets}

OAI \cite{nevitt2006} is a long-term, multicenter observational study on knee osteoarthritis. This public-private partnership aims to develop a research resource for evaluating biomarkers of osteoarthritis as potential surrogate endpoints for disease onset and progression. The study design includes a cohort of 4,796 participants aged 45-79 at recruitment, with high retention rates throughout the study. Data collection involves clinical visits, biospecimen collection, imaging, and questionnaires at various time points. The OAI has made a wealth of data publicly available, including clinical data, images, and biospecimens from visits between February 2004 and October 2015, accessible through the OAI Online website. 

MOST \cite{segal2013} is a prospective observational study focused on risk factors for the development and progression of knee osteoarthritis and knee pain. Sponsored by the National Institutes of Health/National Institute on Aging, MOST is a collaborative effort involving four core sites across the United States. The study design includes multiple data collection points, with clinical visits, telephone interviews, and imaging assessments conducted at baseline, 15, 30, 60, and 84 months. MOST has collected a comprehensive dataset, including clinical data from up to 3,026 participants, radiograph images of up to 6,030 knees, MR images of up to 5,029 knees, and various radiograph and MRI assessments. The study also tracks outcomes such as knee and hip replacements, radiographic OA, and mortality. The MOST dataset was used to test the generalization of the trained models.

The OAI and MOST were approved by the Internal Review Boards at the University of California at San Francisco, Boston University Medical Center, and each individual clinical recruitment site and were performed in compliance with the Health Insurance Portability and Accountability Act. All subjects signed written informed consent.

\subsection{Subject Cohort}
Balanced case-control cohorts were selected from both OAI and MOST sets by matching case subjects and control subjects with respect to baseline demographic variables associated with knee OA progression including age, sex, ethnicity, and body mass index (BMI).  Case subjects were individuals who received a TKR in either knee after baseline enrollment. Control subjects were individuals who had not undergone a TKR in either knee and were present at the latest available follow-up visit (108 months for OAI participants or 84 months for MOST participants). If a patient underwent TKR in both knees during OAI data collection, the knee that first underwent TKR was included. Each case patient with TKR was matched to a control subject without TKR who had the same age, sex, and ethnicity, with an additional constraint on the baseline BMI within a 10\% tolerance. The dataset comprised one knee per participant, using either the left or right knee from each case and control subject in the matched pairs. 

Our analysis was conducted using two distinct datasets: one comprised of radiographs and the other comprised of MRIs. A total of 364 case-control pairs with a baseline radiograph were identified from the 4796 subjects in the OAI database.  Only 354 of these 364 case-control pairs had baseline MRI scans. Subjects were excluded if they had TKR at baseline, received partial TKR over the course of follow-up, had inflammatory arthritis, were missing baseline or 108-month follow-up information, or did not match with a case or control subject.  For the MRI models, the coronal intermediate-weighted turbo spin-echo (COR IW-TSE) sequence in the MRI scans was used for training and validating the models. 

The data from Multicenter Osteoarthritis Study (MOST) was used to test the generalization capabilities of the trained models. Similar to the OAI, two datasets were created: one comprised of radiographs and the other comprised of MRI scans. A total of 574 case-control pairs with baseline radiographs were identified from the 3026 subjects In the MOST database.  Only 378 of these case-control pairs had baseline MRI. The same exclusion criteria as OAI were used here except the last follow-up information was at 84 months instead of 108 months from the baseline. For MOST, High Resolution Coronal Short-Tau Inversion Recovery (HR COR STIR) sequence was used for testing, due to its shared coronal view with the COR IW TSE sequence which was used for training the models. The idea was that maintaining consistency in the imaging plane could possibly help the models generalize better. To address the specific characteristics of each dataset, dedicated models were trained separately for the radiographs and MRI scans. These models were then evaluated on their corresponding test sets to compare model performances. 

\begin{figure}
    \centering
    \begin{subfigure}[b]{0.49\linewidth}
        \centering
        \includegraphics[width=\linewidth]{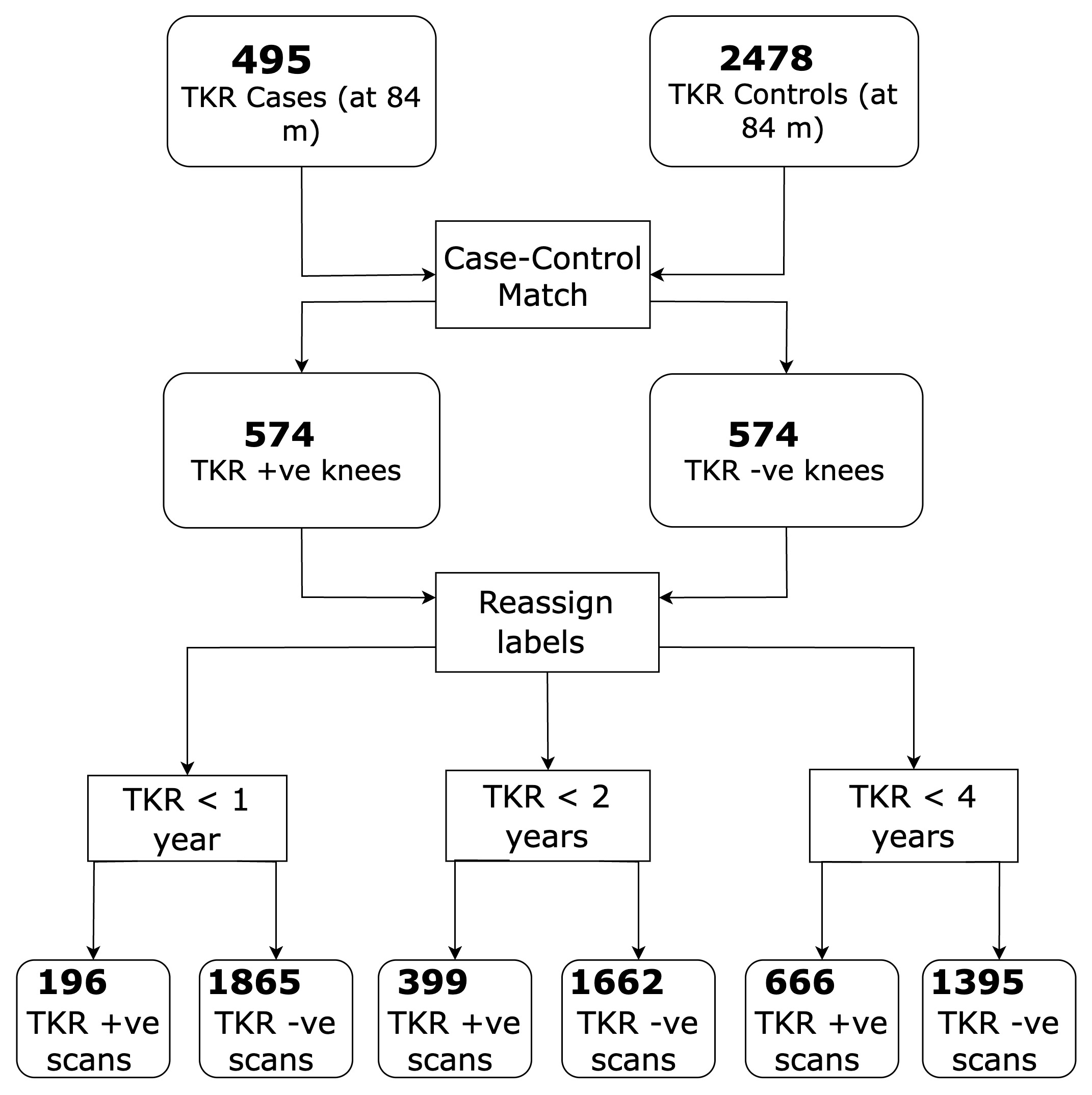}
        \caption{MOST Cohort}
        \label{fig:most_cohort}
    \end{subfigure}
    \hfill
    \begin{subfigure}[b]{0.49\linewidth}
        \centering
        \includegraphics[width=\linewidth]{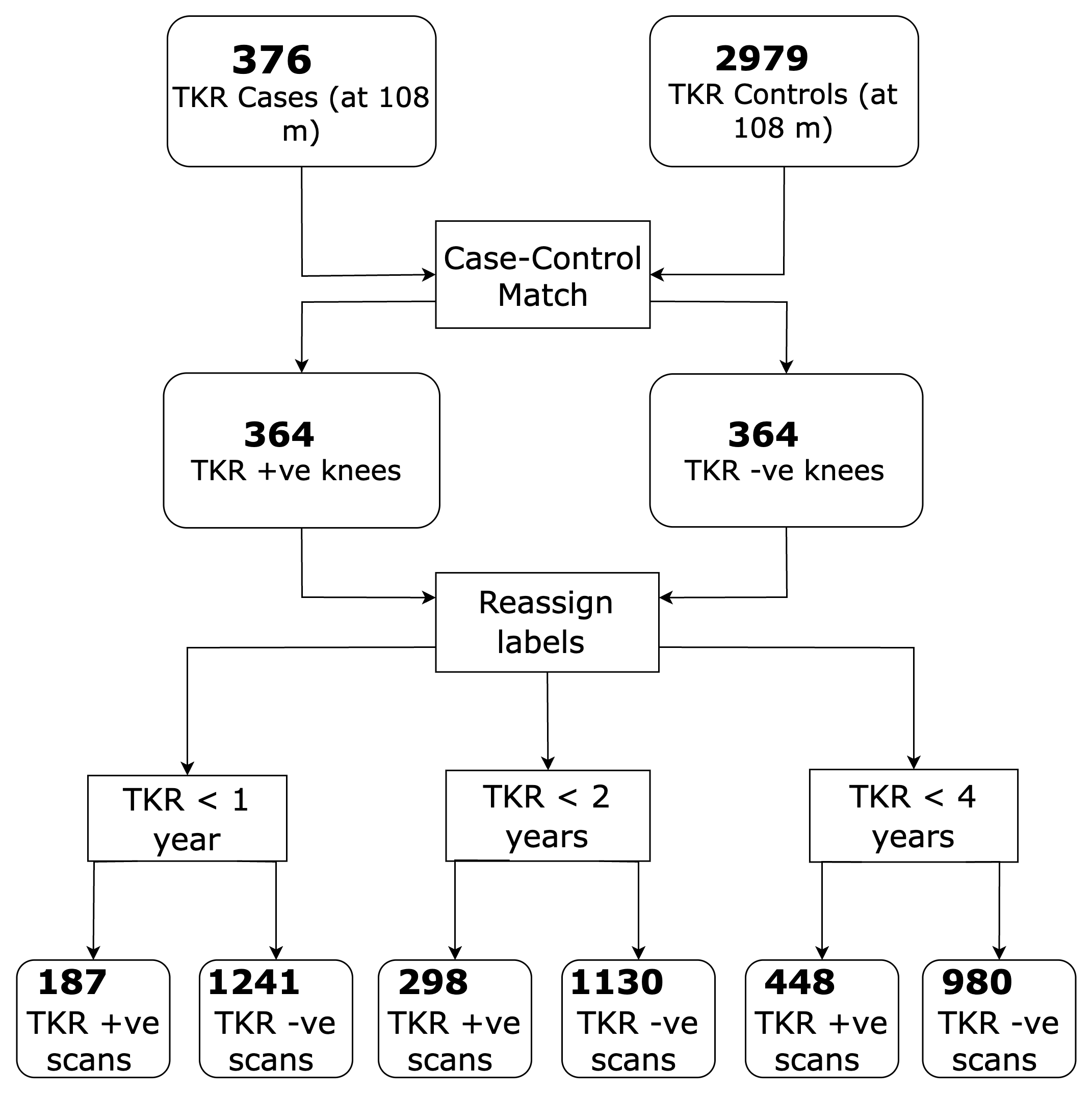}
        \caption{OAI Cohort}
        \label{fig:oai_cohort}
    \end{subfigure}
    \caption{Radiograph Cohort selection pipeline for the MOST and OAI datasets}
    \label{fig:rad_cohort}
\end{figure}

\begin{figure}
    \centering
    \begin{subfigure}[b]{0.49\linewidth}
        \centering
        \includegraphics[width=\linewidth]{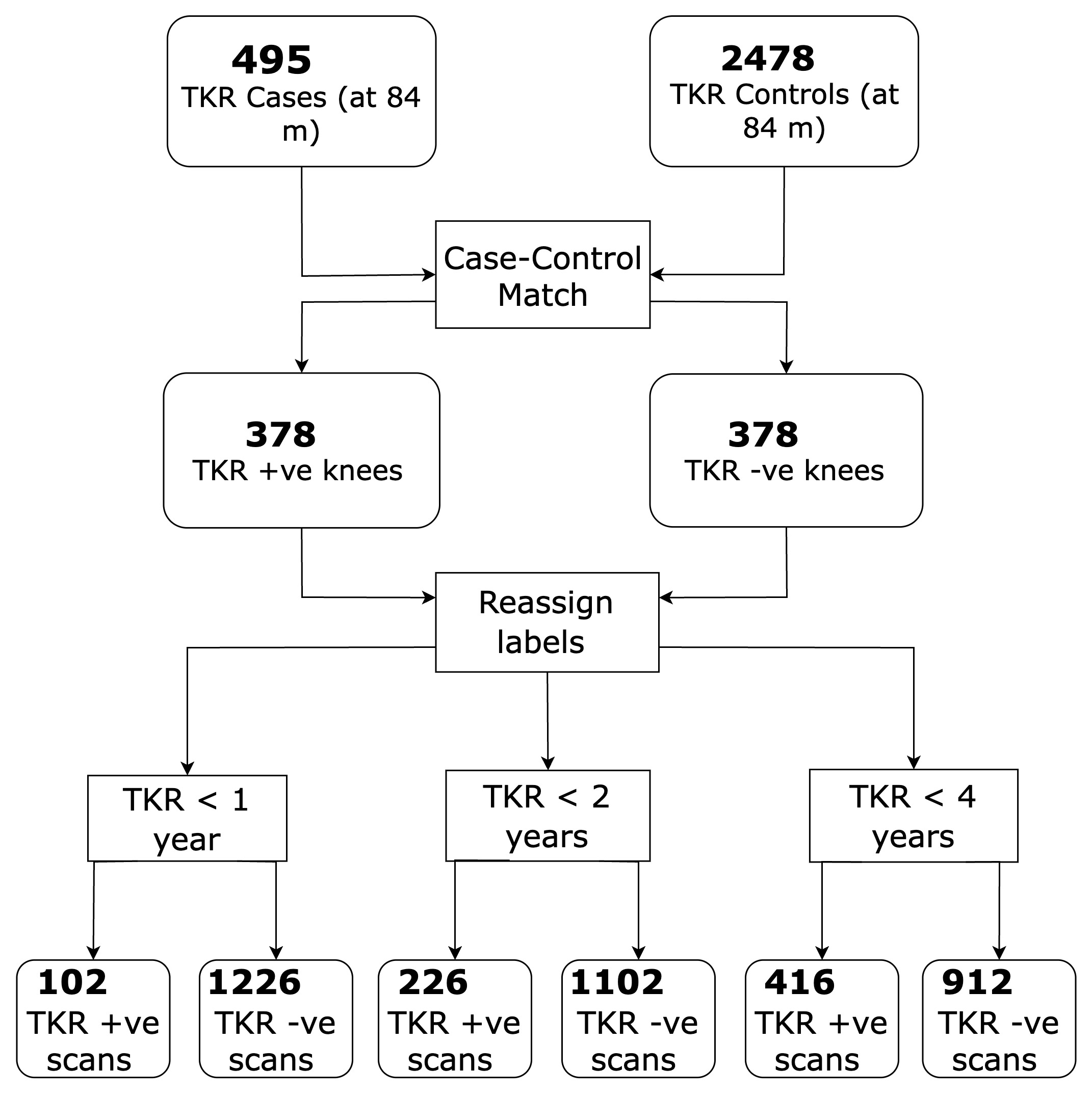}
        \caption{MOST Cohort}
        \label{fig:most_mri_cohort}
    \end{subfigure}
    \hfill
    \begin{subfigure}[b]{0.49\linewidth}
        \centering
        \includegraphics[width=\linewidth]{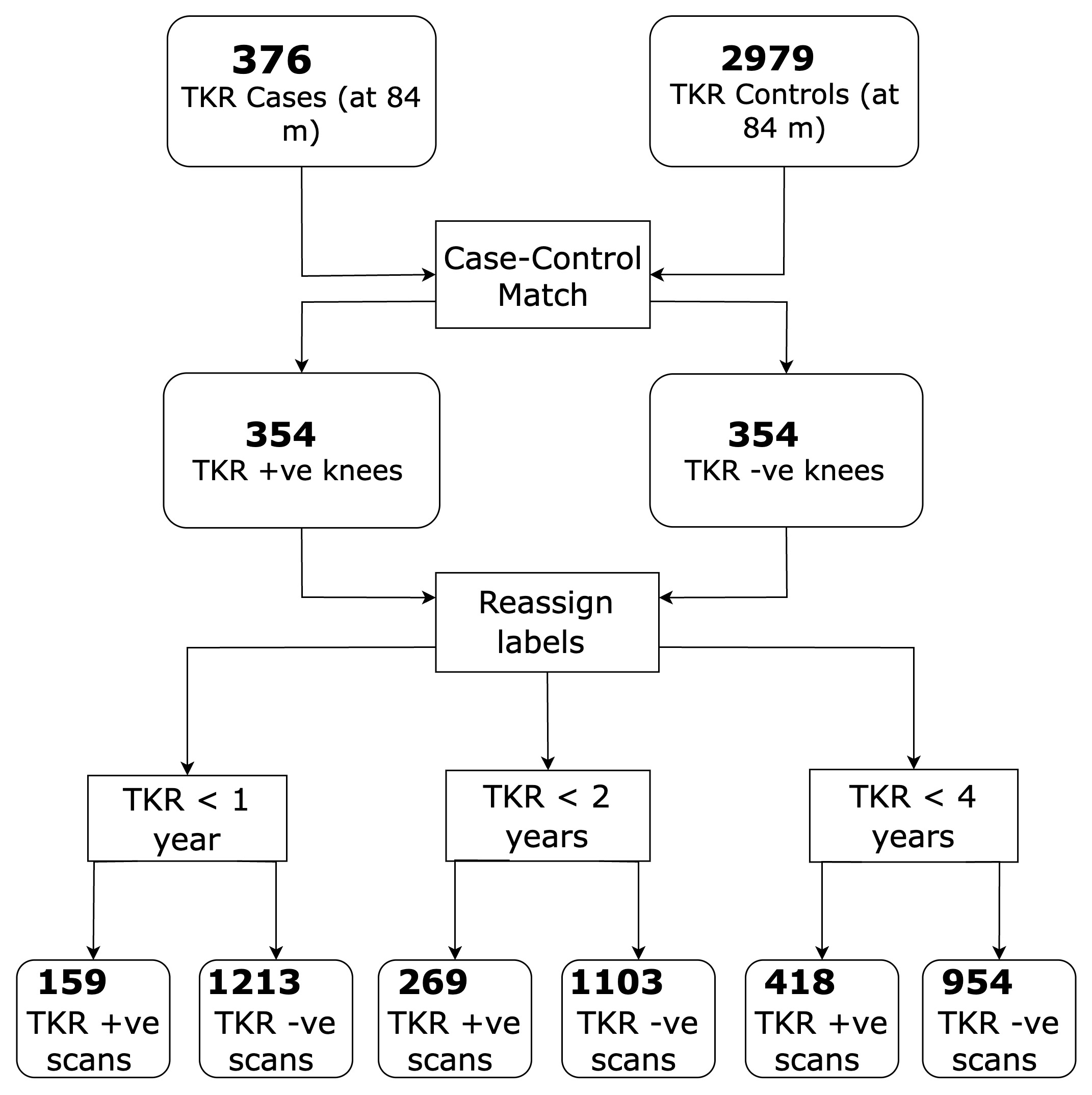}
        \caption{OAI Cohort}
        \label{fig:oai_mri_cohort}
    \end{subfigure}
    \caption{MRI Cohort selection pipeline for the MOST and OAI datasets}
    \label{fig:mri_cohort}
\end{figure}

A summary of the selection of case-control pairs for MRI and radiographs on the OAI and MOST set is illustrated in Figures \ref{fig:rad_cohort} and \ref{fig:mri_cohort}. Table \ref{tab:pat_distribution} summarizes the study cohort characteristics and Table \ref{tab:img_params} provides the imaging parameters of MRI contrasts used in this work.

\subsection{Training and Validation Group}

In both the OAI and MOST datasets, for each subject's knee, 2 scans were used during training (whenever available): the first scan was performed at baseline, and for the second scan, the timing depended on the patient's TKR status. For patients who eventually underwent TKR (TKR-positive), we used their last available scan before the procedure (1-4 years pre-TKR) to capture the most advanced disease stage before surgical intervention. For patients who did not undergo TKR (TKR-negative), we used their last available follow-up scan in the study (with additional availability of 4 year follow up information) to maximize the time period between the two scans. The use of two scans was needed to apply the risk constraint, which assumes that TKR risk on the second scan is always greater than or equal to the TKR risk on the first scan. To increase the training and validation dataset size and scope, knees with only one available radiograph or MRI scan (only the baseline) were also included. When training models on these single-image instances, the methodology was limited to applying the standard cross-entropy loss function, as the risk constraint, which requires a future scan, could not be applied.


The OAI set was split in a stratified manner into 7 folds and nested cross-validation was used in training the models. All splits were performed at the subject level. For nested cross-validation, the dataset was first divided into 7 folds for the outer loop, with each fold serving as a test set once, while the remaining 6 folds were used for training and validation. For each outer fold, the remaining 6 folds underwent another round of 6-fold cross-validation, creating 6 train-validation splits (inner fold) corresponding to each outer fold. In the model training phase, for each inner fold, a model was trained on the training data and validated on the corresponding validation set (effectively using 5/7 of the total data for training, 1/7 for validation, and 1/7 for testing in each complete iteration). This process yielded a total of 42 trained models for each modality (radiograph and MRI). The models were further tested on the entire MOST cohort.

\begin{table}[htbp]
\fontsize{8pt}{9pt}\selectfont 
\centering
\begin{tabular}{@{}p{1.4cm} *{7}{p{1.3cm}} @{}}
\hline
\addlinespace[2.5pt]
\textbf{Modality} & \textbf{Cohort} & \textbf{\# of Patient Knees} & \textbf{\# of Images} & \textbf{\# of males} & \textbf{\# of females} & \textbf{Age (mean $\pm$ std)} & \textbf{BMI (mean $\pm$ std)} \\
\hline
\addlinespace[1.5pt]
\multirow{2}{*}{Radiograph} & OAI & 728 & 1428 & 284 & 444 & 63.6 $\pm$ 8.2 & 29.8 $\pm$ 4.6 \\
 & MOST & 1148 & 2061 & 352 & 796 & 64.3 $\pm$ 7.4 & 31.8 $\pm$ 5.5 \\
 \addlinespace[1.5pt]
\hline
\addlinespace[1.5pt]
\multirow{2}{*}{MRI} & OAI & 708 & 1372 & 276 & 432 & 63.7 $\pm$ 8.2 & 29.7 $\pm$ 4.5 \\
 & MOST & 756 & 1328 & 180 & 576 & 64.9 $\pm$ 7.3 & 30.7 $\pm$ 4.7 \\
 \addlinespace[1.5pt]
\hline
\end{tabular}
\caption{Demographic characteristics of patient cohorts from OAI and MOST datasets. The table presents the distribution of patients by imaging modality (Radiograph and MRI), showing the number of unique knees, total images acquired, gender distribution, and mean age and BMI with standard deviations.}
\label{tab:pat_distribution}
\end{table}

\subsubsection{Prediction Tasks}
\label{pred_tasks}

The models were trained to predict TKR occurrence within three different time horizons from any given scan date:
\begin{itemize}
    \item 1-year TKR prediction – Predicting whether a patient will undergo TKR within 1 year after the scan date
    \item 2-year TKR prediction – Predicting whether a patient will undergo TKR within 2 years after the scan date
    \item 4-year TKR prediction – Predicting whether a patient will undergo TKR within 4 years after the scan date
\end{itemize}

For patients with a single scan, the model can predict a risk normally, whereas for patients with two available scans, predictions were made separately for each scan, with the second scan's prediction incorporating information from the first through our risk constraint (the evaluation setup is described in detail in Section \ref{eval_subsection}). It's important to note that for 1-year TKR prediction, baseline scans (scan 1) were predominantly TKR-negative, as patients who underwent TKR within a year of their baseline visit typically did not have a follow-up scan (scan 2) before their surgery. 

The selection of these time horizons was guided by dataset constraints and clinical relevance. While we had access to up to 9 years of follow-up information from baseline visits, we focused on shorter time horizons since extending past 4 years meant loss of follow-up information for a lot of control patients after the scan 2. For TKR-positive patients, we used their baseline scan and their last available scan prior to surgery, ensuring that our models learned from the progression patterns leading up to TKR.

To effectively understand the motivation for our risk formulation approach, we categorize patients into three distinct groups based on their TKR status across sequential scans:
\begin{itemize}
    \item Progression Group (Set 1): Patients whose status changes from TKR-negative in their first scan to TKR-positive in their second scan. This group represents active disease progression cases.
    \item Stable Negative Group (Set 2): Patients who remain TKR-negative across both scans. This group represents stable cases or slower progression.
    \item Stable Positive Group (Set 3): TKR-positive patients in both scans. Note that this group is empty by design for the 1-year TKR prediction window, as TKR-positive patients in their first scan typically undergo surgery before a second scan can be obtained.
\end{itemize}

Our risk formulation approach primarily aims to nudge the model to better differentiate between the Progression Group (Set 1) and Stable Negative Group (Set 2), learning the subtle imaging patterns that indicate progression toward TKR. However, as the prediction time horizon extends (e.g., to 4 years), we observed a shift in group distributions: the Progression Group (Set 1) decreases in size relative to the Stable Positive Group (Set 3). This distribution shift influenced our selection of prediction horizons (up to 4 years), balancing clinical relevance with the need to maintain sufficient presence of each patient group.
\begin{table}[htbp]
\centering
\fontsize{8}{9}\selectfont
\begin{tabular}{lcccccccc}
\toprule
\textbf{Contrast} & \textbf{TE} & \textbf{TR} & \textbf{TI} & \textbf{FOV} & \textbf{ST} & \textbf{ISR} & \textbf{Matrix Size} & \textbf{Bandwidth} \\
 & \textbf{(ms)} & \textbf{(ms)} & \textbf{(ms)} & \textbf{(mm)} & \textbf{(mm)} & \textbf{(mm$^2$)} & & \textbf{(Hz/pixel)} \\
\midrule
COR IW-TSE & 20 & 3700 & NA & 140 & 3.0 & 0.36$\times$0.36 & 384$\times$384 & 352 \\
\addlinespace[1.5pt]
COR STIR & 35 & 4800 & 100 & 140 & 3.0 & 0.55$\times$0.72 & 256$\times$192 & NA \\
\bottomrule
\end{tabular}
\caption{Imaging parameters of the coronal T1-weighted turbo spin-echo (COR IW-TSE) sequences performed in the MRI examination of subjects in the OAI database, and the coronal short-tau inversion recovery (COR STIR) sequences performed in the MRI examination of subjects in the MOST database.}
\label{tab:img_params}
\end{table}

\subsection{Approach - Modified Risk Formulation}

To strictly enforce the condition that TKR risk increases or remains the same over time, the standard risk formulation (where risk is the sigmoid of logit output from the model) was modified in the following way (RiskFORM1):

$$p(E|x_1) = \hat{y}_1 = \sigma(f(x_1))$$

$$p(E|x_2, x_1) = \hat{y}_2 = 1 - p(E'|x_1) * (1 - \sigma(f(x_2))) = 1 - (1 - \sigma(f(x_1))) * (1 - \sigma(f(x_2)))$$

where,
\begin{itemize}
\item $E$: The event that a TKR is needed within the prediction timeframe  t = 1, 2, or 4 years. $E'$ implies TKR is not needed. 
\item $x_1$: First scan (radiograph or MRI) of a patient
\item $x_2$: \text{Second scan of the same patient (acquired later)}
\item $\hat{y}_1$: \text{Predicted TKR risk for first scan}
\item $\hat{y}_2$: \text{Predicted TKR risk for second scan}
\item $\sigma(z)$: \text{Sigmoid function that converts scores to probabilities: } $\sigma(z) = 1/(1 + e^{-z})$
\item $f(x)$: \text{Deep learning model that produces a real-valued score}
\end{itemize}
For Scan 1 ($x_1$), the prediction happens normally i.e sigmoid of the logit. Note that with this formulation, for any  $f(x_1), f(x_2) \in \mathbb{R}$, provides the desired property that $\hat{y}_1 \leq \hat{y}_2$. 

\text{The proof is simply as follows: since } $(1 - \sigma(f(x_2))) \in (0,1)$

$$p(E'|x_1) * (1 - \sigma(f(x_2))) \leq p(E'|x_1)$$

\text{which means,}

$$1 - p(E'|x_1) * (1 - \sigma(f(x_2))) \geq 1 - p(E'|x_1)$$

$$=> p(E|x_2, x_1) \geq p(E|x_1)$$

Additionally, instead of using a single model $f$ for processing both scans, two separate models $f,g$ can be trained (RiskFORM2) and the TKR risks for this formulation is given by,

$$p(E|x_1) = \hat{y}_1 = \sigma(f(x_1))$$

$$p(E|x_2, x_1) = \hat{y}_2 = 1 - p(E'|x_1) * \sigma(g(x_2)) = 1 - (1 - \sigma(f(x_1))) * \sigma(g(x_2))$$

where $g(x)$ is the second deep learning model used exclusively for processing scan 2.

The final loss function for both formulations is:

$$L_{total} = BCE(y_1, \hat{y}_1) + BCE(y_2, \hat{y}_2)$$

where $y_1$, $y_2$ are True TKR status for the first and second scans respectively. In the case of a knee having a single radiograph or MRI scan, the TKR risk is $\hat{y}_1$  and the loss  $L_{total} = BCE(y_1, \hat{y}_1)$ is applied during training. In RiskFORM2, both models $f,g$ will be trained concurrently whenever two scans of a knee are available.

Note that in RiskFORM1, we use $(1-\sigma(f(x_2)))$ as a multiplier because this ensures that as $f(x_2)$ increases (indicating higher risk), the value of $(1-\sigma(f(x_2)))$ decreases, which in turn reduces the probability of no TKR event, thereby increasing $p(E|x_2,x_1)$. This creates the desired positive correlation between the model output $f(x_2)$ and the final risk prediction.  This formulation is necessary because it maintains the integrity of the risk prediction for scan 1, where $p(E|x_1) = \hat{y}_1 = \sigma(f(x_1))$, so $f(x)$ should always be positively correlated to TKR risk for correct evaluation.

\begin{figure}
    \centering
    \begin{subfigure}[b]{0.67\textwidth}
        \centering
        \includegraphics[width=\textwidth]{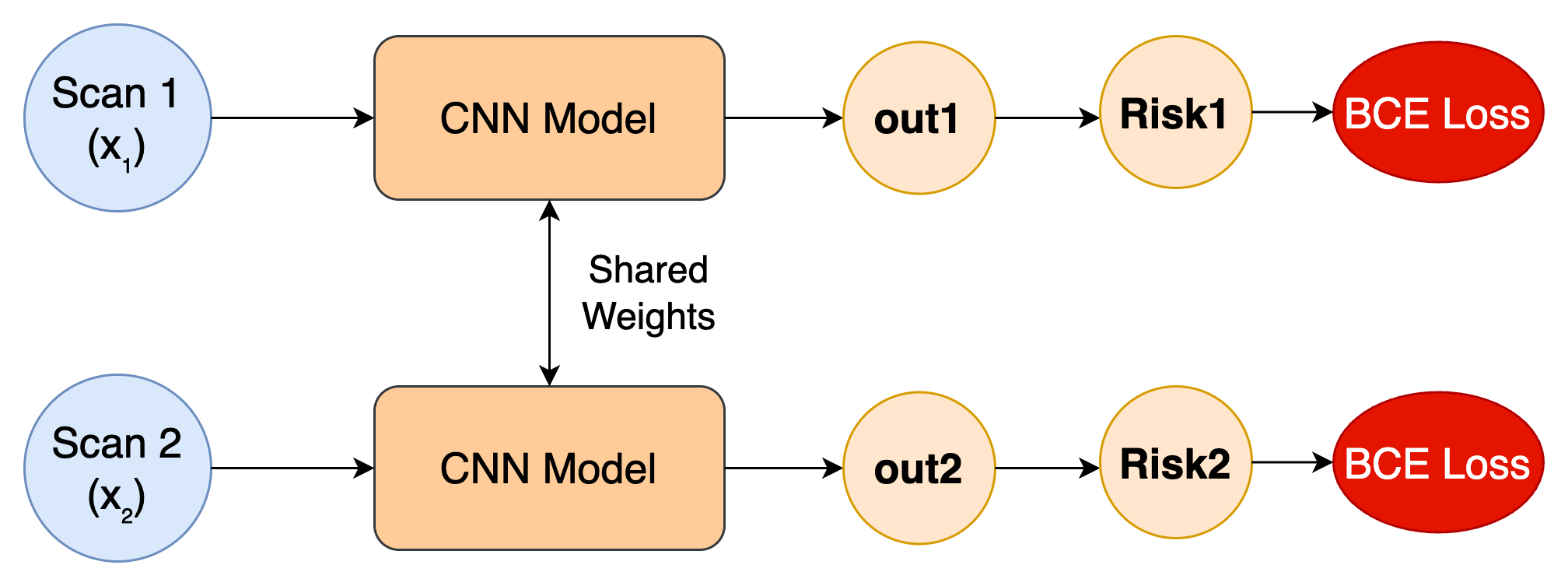}
        \caption{}
        \label{fig:baseline}
    \end{subfigure}
    
    \vspace{1cm}
    
    \begin{subfigure}[b]{0.67\textwidth}
        \centering
        \includegraphics[width=\textwidth]{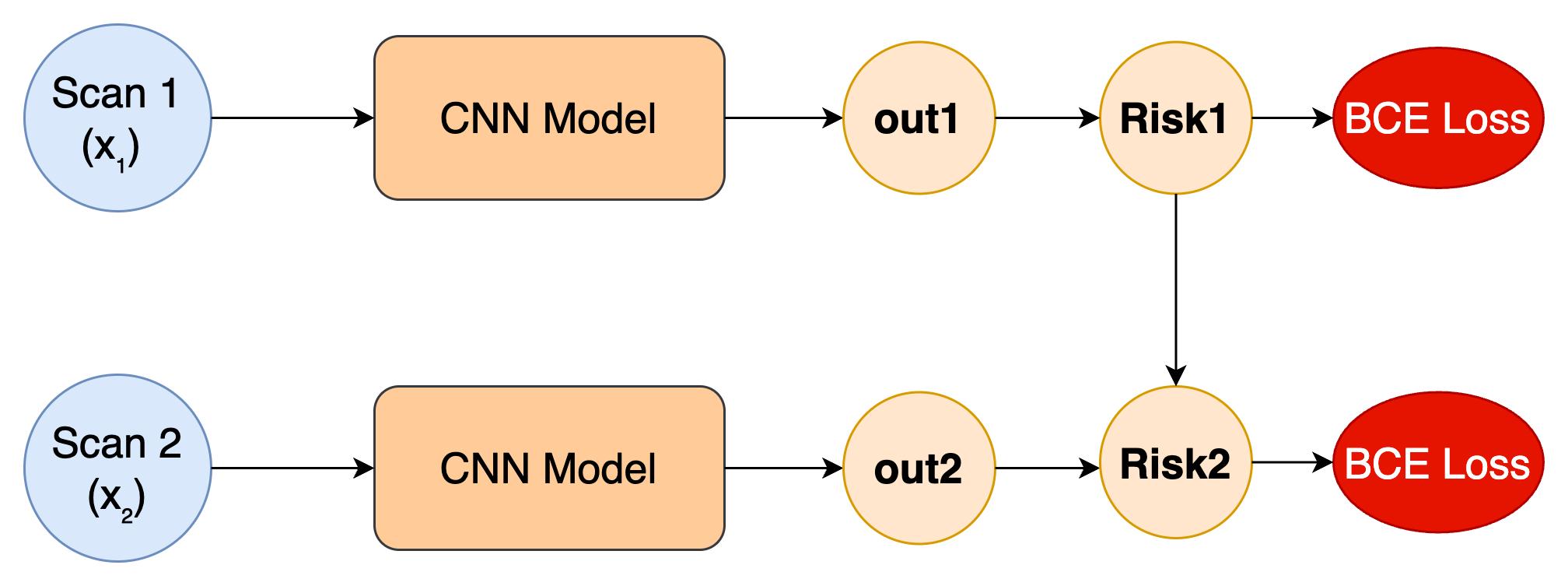}
        \caption{}
        \label{fig:riskform}
    \end{subfigure}
    \caption{(a) The Baseline training approach without any risk constraint. (b) The proposed modified risk formulation scheme that effectively conditions on past scan meaning it utilizes the predicted risk (Risk1) for Scan 1 explicitly to compute risk for the Scan 2 (Risk 2).}
    \label{fig:riskform_vis}
\end{figure}

In contrast, for RiskFORM2, since we use a separate model $g(x_2)$ for the second scan, we can directly use $\sigma(g(x_2))$ as the multiplier. Here, $g(x_2)$ will produce scores that are inversely related to TKR risk. Figure \ref{fig:riskform_vis} illustrates the proposed approaches.

\subsection{Comparative Modeling Approaches}
To evaluate the effectiveness of our proposed risk constraint methods, we experimented with several comparative approaches:

\textbf{Baseline Approach}: We trained separate models for radiographs and MRI scans using standard binary cross-entropy loss without any progressive risk constraints. These models were trained on the complete dataset, including both first and available second scans, treating longitudinal scans of the same patient as independent datapoints.

 \textbf{Risk Constraint Via Regularization}: We applied two regularization techniques to enforce the risk constraint. First, we implemented Contrastive Regularization (ConReg) based on Li et al. [35], who successfully encoded OA severity over time using contrastive loss with Siamese architecture. Second, we applied Risk Regularization (RiskReg), directly applying the risk constraint to model outputs using a margin loss \cite{weston1999}, as shown in Figure \ref{fig:reg_app_vis}. A more detailed explanation of these regularization approaches is provided in the Appendix.

\begin{figure}
    \centering
    \begin{subfigure}[b]{0.75\textwidth}
        \centering
        \includegraphics[width=\textwidth]{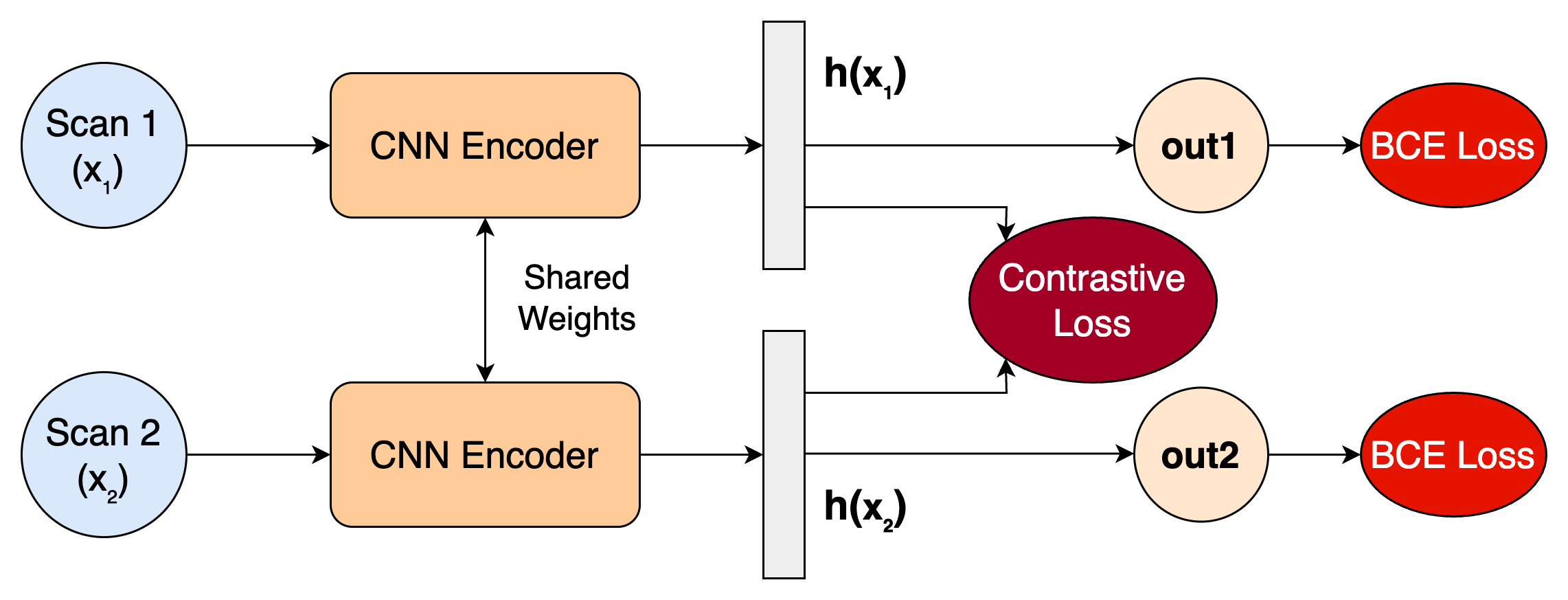}
        \caption{}
        \label{fig:ConReg}
    \end{subfigure}
    
    \vspace{1cm}
    
    \begin{subfigure}[b]{0.67\textwidth}
        \centering
        \includegraphics[width=\textwidth]{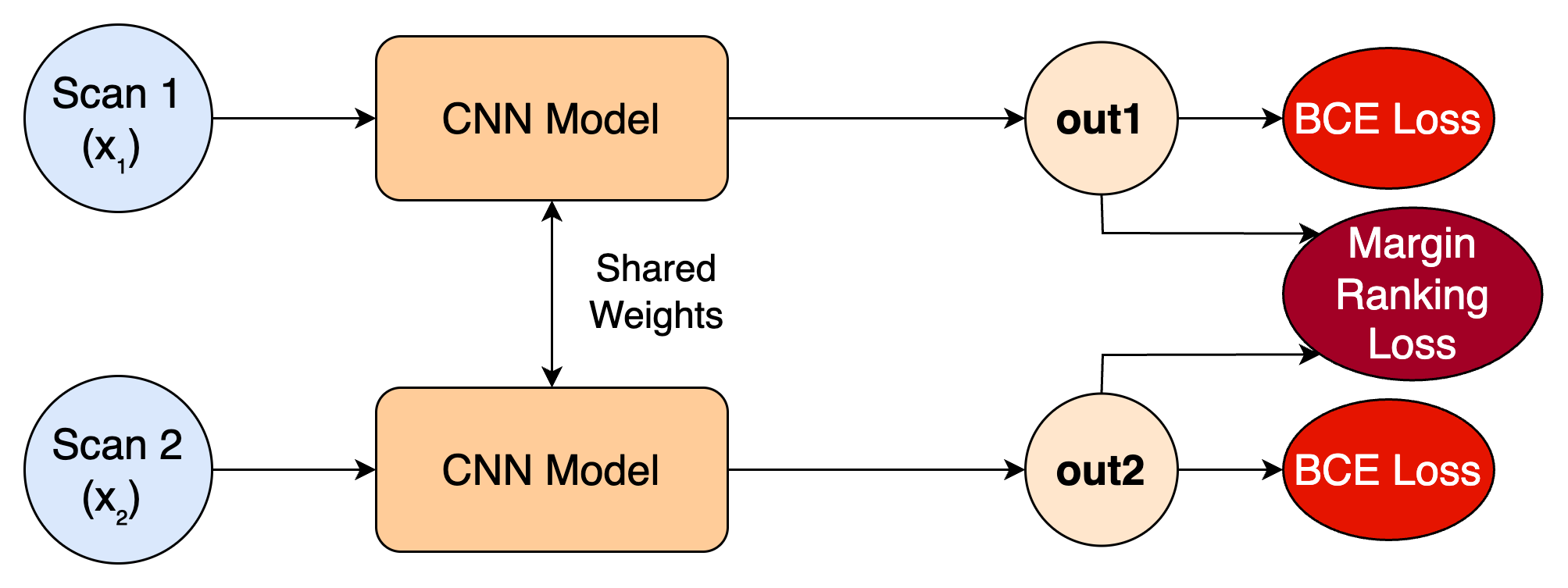}
        \caption{}
        \label{fig:riskreg}
    \end{subfigure}
    \caption{Regularization methods for enforcing the risk constraint (a) ConReg via Siamese loss and (b) RiskReg via margin loss.}
    \label{fig:reg_app_vis}
\end{figure}

\subsection{Model Training}
The radiograph and MRI models were trained to predict TKR within 1-year, 2-years, and 4-years using scans from OAI dataset. For the radiograph models, Resnet34 architecture \cite{he2016} was used with ImageNet pre-trained initialization. For the MRI models, 3D resnet18 with Kaiming initialization \cite{he2015} was used along with Adam optimizer \cite{kingma2014}. For training the RiskReg models, a margin of 2.0 and gamma 1.0 were chosen based on preliminary hyperparameter tuning experiments.

For evaluation on the OAI cohort (training cohort), each of the 7 outer folds was evaluated using an ensemble of its corresponding 6 trained models, with performance metrics computed over the entire cohort by combining results from all 7 folds. In the evaluation on the MOST cohort (external validation), all 42 trained models were ensembled to make predictions, and performance metrics were computed over the entire MOST cohort. For model evaluation, the Area Under the Receiver Operator Curve (AUROC) and the Area Under the Precision-Recall Curve (AUPRC) metrics were computed.

To balance computational efficiency with model performance, a focused approach to hyperparameter tuning was adopted. Hyperparameter optimization was performed using only a single subset of the data, corresponding to the first inner fold of the first outer fold in the nested cross-validation split. Key hyperparameters optimized included learning rate, weight decay, and constraint-specific hyperparameters. The manual optimization process involved conducting initial experiments to establish a baseline, independently adjusting the learning rate and weight decay, and fine-tuning constraint-specific hyperparameters based on the model's requirements.

The effectiveness of each hyperparameter configuration was assessed based on improved performance metrics on the validation set, with the best configuration chosen based on its ability to enhance model performance on the validation data. The optimized hyperparameters were then applied consistently across all folds in the nested cross-validation. This approach assumes that the optimal hyperparameters found for this subset will generalize well to the entire dataset and significantly reduce computational time compared to performing hyperparameter tuning for each fold.

\subsection{Evaluation Framework}
\label{eval_subsection}

The model performance was evaluated at the individual scan level rather than the patient level. For patients with two available scans:
\begin{itemize}
    \item The model generates an independent prediction for the first scan ($p(E|x_1)$).
    \item Using our risk constraint approach, it generates a prediction for the second scan that accounts for the first scan's prediction.
    \item Both predictions are evaluated independently when computing performance metrics
\end{itemize}

This scan-level evaluation approach provides a more comprehensive assessment of the model's performance while ensuring fair comparison between the different approaches. Notably, this evaluation strategy prevents artificial performance inflation that could occur if the risk constraint models simply assigned higher risks to all second scans. By evaluating each scan independently against its respective ground truth label, we ensure that improvements in performance metrics reflect genuine predictive ability rather than systematic bias. For example, if a patient has both a baseline scan and a follow-up scan, each scan contributes separately to the evaluation metrics, requiring accurate predictions at both timepoints rather than just capturing the relative ordering between them.

To understand the significance of performance differences between models, 95\% confidence intervals for both AUROC and AUPRC metrics were used. Further, statistical significance of improvements in diagnostic performance was analyzed by comparing AUROC differences between Baseline and RiskFORM2 models using the Delong test \cite{delong1988}.

\subsection{Analysis of Results across Patient Subgroups}

The study further examined performance metrics within different subsets of patients for TKR prediction within 4 years on the OAI radiograph testset (4 years was picked because it's the longest time horizon in this work and patients with single scan were not used in this analysis). While the prediction tasks in Section \ref{pred_tasks} categorized patients into three sets based on their TKR progression status, for our performance analysis, we created four distinct analytical cohorts to comprehensively evaluate our models:
\begin{itemize}
    \item Cohort 1 (TKR progression): All scans (both scan 1s and scan 2s) of patients who were TKR negative in scan 1 and progressed to TKR positive in scan 2 (equivalent to Set 1 in Section
    \ref{pred_tasks}).
    \item Cohort 2 (TKR Progression + Stable Negatives): All scans of patients who were TKR negative in scan 1 and progressed to TKR positive in scan 2, alongside patients who remained TKR negative in both scans 1 and 2 (combining Set 1 and Set 2).
    \item Cohort 3: All scans of patients who were TKR negative in scan 1 and progressed to TKR positive in scan 2, alongside patients who remained TKR positive in both scans 1 and 2 (combining Set 1 and Set 3).
    \item Cohort 4: All scans of patients who remained TKR negative in both scans 1 and 2, alongside patients who remained TKR positive in both scans 1 and 2 (combining Set 2 and Set 3).
\end{itemize}
	
These analytical cohorts were designed to isolate different aspects of model performance, particularly how well our approaches identified the progression cases versus stable cases.

To further evaluate the discriminative capabilities of the models, an in-depth analysis was conducted to assess performance within each KLG. This analysis aimed to ensure that the models' predictions were not simply correlated with the KL grades but offered additional predictive value. AUROC metrics were again computed for the task of predicting TKR within 4 years, using the OAI radiograph testset.

The dataset was stratified by KLG (ranging from 0 to 4), and AUROC scores were calculated separately for scans within each grade. This approach allowed for a nuanced understanding of how the models performed across different severities of osteoarthritis, as indicated by the KL grading system. By examining the models' performance within each KLG, we could ascertain whether they provided meaningful predictions beyond what could be inferred from the KL grade alone. This analysis is crucial for validating the models' clinical utility, as it demonstrates their ability to differentiate between patients who will and will not require TKR within the same KL grade cohort.

\section{RESULTS}


From Tables \ref{tab:tkr_prediction_rad} and \ref{tab:tkr_prediction_mri}, the modified risk formulation approaches demonstrated superior performance compared to both the baseline and regularization approaches across radiographs and MRIs. Notably, RiskFORM2 achieved the best overall performance on all TKR prediction timeframes (1-year, 2-year, and 4-year), while RiskFORM1 consistently outperformed the baseline approach. These results indicate that the proposed risk formulation methods enhance predictive capability for TKR across different imaging modalities and prediction horizons.

\begin{table}[ht]
\centering
\fontsize{6.5pt}{8pt}\selectfont 
\begin{tabular}{@{}p{1.1cm} *{12}{p{0.7cm}} @{}}
\toprule
\multirow{3}{=}{\textbf{Approach}} & \multicolumn{4}{c}{\textbf{1 year TKR}} & \multicolumn{4}{c}{\textbf{2 years TKR}} & \multicolumn{4}{c}{\textbf{4 years TKR}} \\ 
\cmidrule(lr){2-5} \cmidrule(lr){6-9} \cmidrule(lr){10-13}
 & \multicolumn{2}{c}{\textbf{OAI}} & \multicolumn{2}{c}{\textbf{MOST}} & \multicolumn{2}{c}{\textbf{OAI}} & \multicolumn{2}{c}{\textbf{MOST}} & \multicolumn{2}{c}{\textbf{OAI}} & \multicolumn{2}{c}{\textbf{MOST}} \\
\cmidrule(lr){2-3} \cmidrule(lr){4-5} \cmidrule(lr){6-7} \cmidrule(lr){8-9} \cmidrule(lr){10-11} \cmidrule(lr){12-13}
 & {\textbf{AUC}} & {\textbf{AUPRC}} & {\textbf{AUC}} & {\textbf{AUPRC}} & {\textbf{AUC}} & {\textbf{AUPRC}} & {\textbf{AUC}} & {\textbf{AUPRC}} & {\textbf{AUC}} & {\textbf{AUPRC}} & {\textbf{AUC}} & {\textbf{AUPRC}} \\
 & {\textbf{(95\% CI)}} & {\textbf{(95\% CI)}} & {\textbf{(95\% CI)}} & {\textbf{(95\% CI)}} & {\textbf{(95\% CI)}} & {\textbf{(95\% CI)}} & {\textbf{(95\% CI)}} & {\textbf{(95\% CI)}} & {\textbf{(95\% CI)}} & {\textbf{(95\% CI)}} & {\textbf{(95\% CI)}} & {\textbf{(95\% CI)}} \\
\midrule
Baseline & 0.79\newline(0.76,\newline0.83) & 0.34\newline(0.29,\newline0.41) & 0.71\newline(0.68,\newline0.75) & 0.19\newline(0.16,\newline0.24) & 0.83\newline(0.81,\newline0.86) & 0.54\newline(0.49,\newline0.61) & 0.75\newline(0.73,\newline0.78) & 0.39\newline(0.35,\newline0.44) & 0.85\newline(0.82,\newline0.87) & 0.70\newline(0.65,\newline0.75) & 0.79\newline(0.77,\newline0.81) & 0.59\newline(0.55,\newline0.63) \\
\addlinespace 
RiskReg & 0.79\newline(0.76,\newline0.83) & 0.34\newline(0.29,\newline0.40) & 0.72\newline(0.69,\newline0.76) & 0.18\newline(0.15,\newline0.21) & 0.81\newline(0.79,\newline0.84) & 0.52\newline(0.46,\newline0.58) & 0.76\newline(0.73,\newline0.78) & 0.38\newline(0.34,\newline0.43) & 0.84\newline(0.81,\newline0.86) & 0.68\newline(0.63,\newline0.73) & 0.80\newline(0.78,\newline0.82) & 0.60\newline(0.56,\newline0.64) \\
\addlinespace
ConReg & 0.83\newline(0.80,\newline0.86) & 0.38\newline(0.33,\newline0.45) & 0.76\newline(0.73,\newline0.79) & 0.20\newline(0.17,\newline0.25) & 0.83\newline(0.81,\newline0.86) & 0.55\newline(0.49,\newline0.61) & 0.77\newline(0.75,\newline0.79) & 0.39\newline(0.35,\newline0.44) & 0.85\newline(0.83,\newline0.87) & 0.70\newline(0.65,\newline0.74) & 0.80\newline(0.78,\newline0.82) & 0.60\newline(0.56,\newline0.64) \\
\addlinespace
ConReg + \newline RiskReg & 0.81\newline(0.78,\newline0.84) & 0.35\newline(0.30,\newline0.42) & 0.73\newline(0.70,\newline0.76) & 0.19\newline(0.16,\newline0.24) & 0.83\newline(0.80,\newline0.85) & 0.52\newline(0.46,\newline0.58) & 0.76\newline(0.74,\newline0.78) & 0.39\newline(0.34,\newline0.44) & 0.84\newline(0.82,\newline0.86) & 0.70\newline(0.65,\newline0.74) & 0.81\newline(0.79,\newline0.83) & 0.62\newline(0.58,\newline0.66) \\
\addlinespace
RiskFORM2 & \textbf{0.87}\newline(0.84,\newline0.89) & \textbf{0.47}\newline(0.40,\newline0.55) & 0.77\newline(0.74,\newline0.81) & \textbf{0.25}\newline(0.20,\newline0.30) & \textbf{0.85}\newline(0.83,\newline0.88) & \textbf{0.63}\newline(0.57,\newline0.70) & \textbf{0.79}\newline(0.77,\newline0.82) & \textbf{0.43}\newline(0.38,\newline0.48) & \textbf{0.86}\newline(0.84,\newline0.88) & \textbf{0.75}\newline(0.71,\newline0.80) & \textbf{0.81}\newline(0.79,\newline0.83) & \textbf{0.63}\newline(0.59,\newline0.67) \\
\addlinespace
RiskFORM1 & 0.80\newline(0.77,\newline0.83) & 0.39\newline(0.33,\newline0.47) & \textbf{0.78}\newline(0.75,\newline0.81) & 0.23\newline(0.19,\newline0.28) & 0.83\newline(0.80,\newline0.85) & 0.58\newline(0.52,\newline0.64) & 0.78\newline(0.75,\newline0.80) & 0.41\newline(0.36,\newline0.46) & 0.85\newline(0.82,\newline0.87) & 0.72\newline(0.68,\newline0.77) & 0.81\newline(0.79,\newline0.83) & 0.63\newline(0.59,\newline0.67) \\
\bottomrule
\end{tabular}
\caption{Results comparing the Modified Risk formulation approaches to risk regularization and baseline approaches on the OAI and MOST radiograph test sets.}
\label{tab:tkr_prediction_rad}
\end{table}

RiskFORM2 also demonstrated superior generalization capability on the external MOST radiograph test set, despite experiencing a performance decline compared to the OAI testset. Both RiskFORM2 and RiskFORM1 consistently outperformed the baseline approach across all prediction timeframes on the MOST dataset. Notably, while the improvement margin of RiskFORM2 over the baseline decreased as the prediction timeframe extended in the OAI test set, it maintained considerable improvement across all timeframes in the external MOST dataset.
Similar performance patterns were observed with MRI data, where RiskFORM2 generally achieved the best results on both the COR IW TSE sequence in OAI and the COR STIR sequence in MOST. An exception was noted in the 4-year TKR prediction on the OAI test set, where RiskFORM1 slightly outperformed RiskFORM2. As with the radiograph results, the performance advantage of RiskFORM2 over the baseline diminished with increasing prediction timeframes in OAI, while maintaining considerable improvement across all prediction horizons in the MOST dataset.

\begin{table}[ht]
\centering
\fontsize{6.5pt}{8pt}\selectfont 
\begin{tabular}{@{}p{1.1cm} *{12}{p{0.7cm}} @{}}
\toprule
\multirow{3}{=}{\textbf{Approach}} & \multicolumn{4}{c}{\textbf{1 year TKR}} & \multicolumn{4}{c}{\textbf{2 years TKR}} & \multicolumn{4}{c}{\textbf{4 years TKR}} \\ 
\cmidrule(lr){2-5} \cmidrule(lr){6-9} \cmidrule(lr){10-13}
 & \multicolumn{2}{c}{\textbf{OAI}} & \multicolumn{2}{c}{\textbf{MOST}} & \multicolumn{2}{c}{\textbf{OAI}} & \multicolumn{2}{c}{\textbf{MOST}} & \multicolumn{2}{c}{\textbf{OAI}} & \multicolumn{2}{c}{\textbf{MOST}} \\
\cmidrule(lr){2-3} \cmidrule(lr){4-5} \cmidrule(lr){6-7} \cmidrule(lr){8-9} \cmidrule(lr){10-11} \cmidrule(lr){12-13}
 & {\textbf{AUC}} & {\textbf{AUPRC}} & {\textbf{AUC}} & {\textbf{AUPRC}} & {\textbf{AUC}} & {\textbf{AUPRC}} & {\textbf{AUC}} & {\textbf{AUPRC}} & {\textbf{AUC}} & {\textbf{AUPRC}} & {\textbf{AUC}} & {\textbf{AUPRC}} \\
 & {\textbf{(95\% CI)}} & {\textbf{(95\% CI)}} & {\textbf{(95\% CI)}} & {\textbf{(95\% CI)}} & {\textbf{(95\% CI)}} & {\textbf{(95\% CI)}} & {\textbf{(95\% CI)}} & {\textbf{(95\% CI)}} & {\textbf{(95\% CI)}} & {\textbf{(95\% CI)}} & {\textbf{(95\% CI)}} & {\textbf{(95\% CI)}} \\
\midrule
Baseline & 0.79\newline(0.76,\newline0.83) & 0.34\newline(0.28,\newline0.42) & 0.70\newline(0.65,\newline0.75) & 0.17\newline(0.12,\newline0.23) & 0.82\newline(0.79,\newline0.85) & 0.52\newline(0.46,\newline0.60) & 0.66\newline(0.62,\newline0.69) & 0.24\newline(0.21,\newline0.29) & 0.84\newline(0.82,\newline0.86) & 0.71\newline(0.66,\newline0.75) & 0.58\newline(0.55,\newline0.61) & 0.35\newline(0.32,\newline0.39) \\
\addlinespace 
RiskReg & 0.80\newline(0.76,\newline0.83) & 0.34\newline(0.28,\newline0.42) & 0.68\newline(0.63,\newline0.72) & 0.14\newline(0.11,\newline0.20) & 0.83\newline(0.80,\newline0.86) & 0.54\newline(0.48,\newline0.61) & 0.62\newline(0.58,\newline0.66) & 0.21\newline(0.18,\newline0.25) & 0.84\newline(0.82,\newline0.86) & 0.71\newline(0.66,\newline0.75) & 0.59\newline(0.56,\newline0.62) & 0.35\newline(0.32,\newline0.39) \\
\addlinespace
ConReg & 0.78\newline(0.74,\newline0.81) & 0.34\newline(0.28,\newline0.42) & 0.70\newline(0.65,\newline0.74) & 0.17\newline(0.12,\newline0.24) & 0.83\newline(0.80,\newline0.86) & 0.54\newline(0.47,\newline0.61) & 0.60\newline(0.56,\newline0.64) & 0.21\newline(0.18,\newline0.24) & 0.85\newline(0.83,\newline0.87) & 0.71\newline(0.66,\newline0.75) & 0.58\newline(0.55,\newline0.62) & 0.36\newline(0.33,\newline0.40) \\
\addlinespace
ConReg +\newline RiskReg & 0.81\newline(0.78,\newline0.84) & 0.38\newline(0.32,\newline0.46) & 0.70\newline(0.65,\newline0.75) & 0.16\newline(0.12,\newline0.22) & 0.84\newline(0.81,\newline0.86) & 0.55\newline(0.48,\newline0.62) & 0.62\newline(0.59,\newline0.66) & 0.22\newline(0.19,\newline0.26) & 0.84\newline(0.81,\newline0.86) & 0.70\newline(0.65,\newline0.74) & 0.56\newline(0.53,\newline0.59) & 0.34\newline(0.31,\newline0.38) \\
\addlinespace
RiskFORM2 & \textbf{0.84}\newline(0.80,\newline0.87) & \textbf{0.43}\newline(0.35,\newline0.51) & \textbf{0.72}\newline(0.67,\newline0.77) & \textbf{0.19}\newline(0.14,\newline0.27) & \textbf{0.84}\newline(0.81,\newline0.87) & \textbf{0.59}\newline(0.53,\newline0.66) & 0.66\newline(0.62,\newline0.69) & \textbf{0.27}\newline(0.23,\newline0.32) & 0.84\newline(0.82,\newline0.86) & 0.72\newline(0.68,\newline0.77) & \textbf{0.59}\newline(0.56,\newline0.62) & \textbf{0.38}\newline(0.34,\newline0.42) \\
\addlinespace
RiskFORM1 & 0.80\newline(0.77,\newline0.83) & 0.35\newline(0.29,\newline0.43) & 0.69\newline(0.64,\newline0.74) & 0.14\newline(0.11,\newline0.19) & 0.84\newline(0.82,\newline0.87) & 0.57\newline(0.51,\newline0.64) & 0.66\newline(0.63,\newline0.70) & 0.24\newline(0.21,\newline0.29) & \textbf{0.86}\newline(0.83,\newline0.88) & \textbf{0.72}\newline(0.68,\newline0.77) & 0.58\newline(0.55,\newline0.62) & 0.36\newline(0.32,\newline0.40) \\
\bottomrule
\end{tabular}
\caption{Results comparing the Modified Risk formulation approaches to risk regularization and baseline approaches on the OAI (COR IW-TSE) and MOST MRI (COR STIR) test sets.}
\label{tab:tkr_prediction_mri}
\end{table}


Enforcing constraints through regularization approaches (RiskReg and ConReg) generally resulted in poorer performance compared to the baseline approach. While ConReg showed modest effectiveness on the radiograph test set and demonstrated reasonable generalization to the MOST test set, it consistently underperformed the baseline approach when applied to MRI data in both OAI and MOST datasets. Furthermore, RiskReg and the combined RiskReg-ConReg approach exhibited inferior performance to the baseline across all test scenarios, suggesting that these regularization strategies may not be optimal for improving TKR prediction models.

The DeLong test results comparing baseline and RiskFORM 2 approaches revealed evidence of improved predictive performance for shorter-time horizon predictions, particularly for 1-year TKR prediction using radiographs ($p=3.4e-06$) and MRI ($p=0.01$). For 2-year prediction horizons, the evidence was more modest but still suggestive of improvement (radiographs: $p=0.09$, MRI: $p=0.1$). Interestingly, for 4-year predictions, the statistical evidence varied considerably between imaging modalities, with radiographs showing borderline evidence ($p=0.05$) while MRI showed no evidence of difference ($p=0.95$).

For performance within various subcohorts, Table \ref{tab:cohort_comparison} reveals that RiskFORM2 generally outperforms both the baseline and RiskFORM1 across most subcohorts and metrics in predicting TKR status within 4 years. This suggests that the modifications implemented in RiskFORM2 have led to significant improvements in the model's predictive capabilities for TKR outcomes across the board.

In Cohort 1, which focuses purely on TKR progression scans, RiskFORM2 shows a substantial improvement over both Baseline and RiskFORM1 (0.88 compared to 0.79 and 0.78) in discriminating between TKR negative scan 1s and TKR positive scan 2s. The enforced condition ($\hat{y}_1>\hat{y}_2$) likely contributes to this improved performance, as it aligns with the actual progression in this group. 

For Cohort 2, which includes TKR Progression and Stable TKR Negatives, all models perform well in terms of AUROC, with RiskFORM2 outperforming others here too indicating a more subtle understanding of the disease progression.

Cohort 3, comprising TKR Progression and Stable TKR positives, shows marked improvement for RiskFORM2 over other models. The high AUPRC scores indicate good performance in identifying true positives. In contrast, Cohort 4, which includes only Stable TKR positives and negatives, shows relatively similar performance across all models. 

The condition enforcing that risk prediction for scan 2 $>$ prediction of scan 1 in RiskFORM1 and RiskFORM2 appears to be particularly beneficial for cases involving TKR progression (from TKR negative to positive), as expected in Cohorts 1, 2 and 3. This aligns with the clinical expectation of disease progression. 

In terms of precision versus recall, the higher AUPRC scores in Cohorts 1 and 3 for RiskFORM2 indicate improved precision in identifying true positives. This aspect is crucial in clinical settings to reduce unnecessary interventions and to ensure accurate identification of patients likely to require TKR.

\begin{table}[ht]
\centering
\fontsize{9pt}{11pt}\selectfont 
\begin{tabular}{lcccc}
\toprule
\textbf{Patient Group} & \textbf{Metric} & \textbf{Baseline} & \textbf{RiskFORM1} & \textbf{RiskFORM2} \\
\midrule
\multirow{2}{*}{Cohort 1} & AUROC & 0.78 & 0.79 & 0.88 \\
 & AUPRC & 0.76 & 0.8 & 0.9 \\
\midrule
\multirow{2}{*}{Cohort 2} & AUROC & 0.87 & 0.86 & 0.9 \\
 & AUPRC & 0.58 & 0.6 & 0.66 \\
\midrule
\multirow{2}{*}{Cohort 3} & AUROC & 0.73 & 0.76 & 0.83 \\
 & AUPRC & 0.85 & 0.87 & 0.92 \\
\midrule
\multirow{2}{*}{Cohort 4} & AUROC & 0.85 & 0.85 & 0.85 \\
 & AUPRC & 0.63 & 0.65 & 0.65 \\
\bottomrule
\end{tabular}
\caption{Comparison of three approaches (Baseline, RiskFORM1, and RiskFORM2) across different patient subgroups, using AUROC and AUPRC on Radiograph based TKR prediction within 4 years. Cohort 1: patients who were TKR negative in scan 1 and progressed to TKR positive in scan 2. Cohort 2: Cohort 1 plus patients who remained TKR negative in both scans. Cohort 3: Cohort 1 plus patients who remained TKR positive in both scans. Cohort 4: patients who remained TKR negative in both scans, plus those who remained TKR positive in both scans.}
\label{tab:cohort_comparison}
\end{table}
Overall, RiskFORM2 demonstrates superior performance in predicting TKR progression, particularly in cases where progression to TKR occurs. The enforced condition appears to be a key factor in this improvement. These results indicate that RiskFORM2 could be a valuable tool in clinical settings for predicting TKR needs, especially when monitoring patients over time.

Within each KLG, from Table \ref{tab:klg_comparison}, all models exhibit positive discriminative capabilities, with AUROC values consistently exceeding 0.5 across all grades. This indicates a predictive value beyond random chance, regardless of osteoarthritis severity (i.e, models are not overfit to KLG differences). RiskFORM2 demonstrates particularly strong performance for patients with likely or established osteoarthritis (KLG 2-4), achieving AUROC values of 0.73, 0.7, and 0.71, respectively. This suggests a more nuanced understanding of osteoarthritis progression compared to the baseline approach, which achieved lower AUROC values of 0.7, 0.64, and 0.62 for the same KLG levels. RiskFORM1 also outperforms the baseline approach, albeit by a smaller margin, with AUROC values of 0.71, 0.65, and 0.64 for KLG 2-4. These results indicate that both the RiskFORM models, particularly RiskFORM2, offer improved predictive capabilities over the baseline approach, especially in cases of more advanced OA.

\begin{table}[ht]
\centering
\fontsize{9pt}{11pt}\selectfont
\begin{tabular}{lcccc}
\toprule
\textbf{KLG} & \textbf{Baseline} & \textbf{RiskFORM1} & \textbf{RiskFORM2} \\
\midrule
KL 0 & 0.73 & 0.77 & 0.66 \\
\midrule
KL 1 & 0.6 & 0.49 & 0.64 \\
\midrule
KL 2 & 0.7 & 71 & 0.73 \\
\midrule
KL 3 & 0.64 & 0.65 & 0.7\\
\midrule
KL 4  & 0.62 & 0.64 & 0.71 \\
\bottomrule
\end{tabular}
\caption{AUROC Performance Comparison of Baseline, RiskFORM1, and RiskFORM2 Models Across Different KLG for TKR Prediction Within 4 Years. This table demonstrates the models' discriminative capabilities within each KLG, highlighting their ability to predict TKR beyond what might be inferred from the KL grade alone.}
\label{tab:klg_comparison}
\end{table}

\section{DISCUSSION}
This study showed that the modified risk formulation approaches considerably outperformed the baseline approach and regularization approaches for predicting TKR across various time periods. These trends held when models were tested on an external MOST testing set, although with a slight performance decrease. 

The modified risk formulation enhances the performance of TKR prediction models by essentially utilizing longitudinal data to refine its predictions. Unlike the baseline approach, which treats imaging data independently, this approach conditions the risk prediction on a previous scan, utilizing a broader spectrum of subject data to better inform the risk prediction. This additional temporal context provides a more nuanced understanding of disease progression, which is crucial for accurate risk assessment in a progressive disease such as OA. RiskFORM2 utilizes two separate models to independently process image 1 and image 2, while RiskFORM1 employs a single shared model for both images. The dual-model approach of RiskFORM2 allows for greater flexibility in risk adjustment between the two images, leading to improved performance compared to RiskFORM1. Furthermore, the superior performance of the modified risk formulation models on the MOST test set suggests that incorporating the risk constraint improves model generalization and robustness to changes in data distribution. Our study also opens an interesting line of research into developing new risk formulations that can condition on more than one past scan in a scalable and efficient manner.

Interestingly, the relative improvement of RiskFORM1 and RiskFORM2 over the baseline approach decreased as the TKR prediction time-horizon increased in the case of both the OAI and MOST test sets.  However, the modified risk formulation approaches still improved performance for the 4-year TKR prediction, especially for the MOST test set.  This decrease in model performance may be attributed to a reduction in the number of knees available for model training as the number of subjects transitioning from TKR-negative in the first image (i.e., baseline image) to TKR-positive in the second image (i.e., last available follow-up image) decreased as the TKR time horizon increased. However, RiskFORM still demonstrated its utility at the 4-year TKR prediction, showing persistent improvements compared to the baseline model, which typically did not adjust its predictions between two subtly different images. Conversely, the RiskFORM approaches are designed to account for these nuanced differences between the baseline and follow-up images, thereby improving model performance.

In the MRI dataset, similar patterns emerged, with RiskFORM2 generally achieving the highest performance metrics. However, for the 4-year TKR predictions, RiskFORM1 unexpectedly outperformed both the baseline approach and RiskFORM2 on the OAI test set. Given that MRI data is three-dimensional and inherently more complex than radiographs, coupled with a smaller training set, there is an increased risk of overfitting with RiskFORM2, which trains two models, compared to the single model used in RiskFORM1. In contrast, regularization strategies such as RiskReg and ConReg did not improve upon the baseline approach and, in some instances, led to lower performance, particularly when evaluating the MRI scans. This suggests that the added complexity introduced by these approaches may not align well with the underlying data characteristics or model structures. 

Previously, Kim et al. \cite{kim2023} introduced PairNet, a pairwise image ranking network designed to order two medical images from the same patient by identifying which image was captured later. The prediction is based on the difference between the predicted representations from the network's final layer on two image inputs. This serves as a regularization, forcing the model to accurately predict the rank using a cross-entropy loss function, like RiskReg and ConReg.  However, this framework was used exclusively in self-supervised learning and not directly in supervised learning applications. In the realm of supervised learning, the investigators used a similar strategy but only for regression tasks that involve continuous labels such as tumor size, which continuously vary over time - unlike the binary classification in our work. This method was effective for predicting brain age, a task where changes over time are incorporated into the label, unlike in binary classification. To the best of our knowledge, our work is the first to introduce a risk formulation modification to incorporate severity constraints, making it applicable to a wide variety of longitudinal progressive disease datasets, not limited to scenarios with continuously changing labels. 

While direct comparisons with previous studies were challenging due to differences in prediction time horizons, we can provide a rough comparison with work on predicting TKR within a 9-year timeframe. In a previous study, Leung et al \cite{leung2020} utilized ResNet34-based CNN models for predicting knee OA outcomes, including KLGs and 9-year TKR, achieving AUROCs of 0.87 for predicting TKR and over 0.80 for predicting KL grades.

In this work, we observed that TKR prediction performance declined as the follow-up time period shortened from 9 years to 4 years or less. For all approaches—including both the baseline and the modified risk formulation methods—performance metrics typically decreased as the TKR prediction time horizon became shorter. This lower model performance may be attributed to two factors: First, the datasets were not identical. While Leung et al \cite{leung2020} used only baseline radiographs from the OAI dataset for training and testing, our study incorporated additional image data, including sequentially performed radiographs and MRI scans. Secondly, as illustrated by COVID-19 progression prediction models from Shamout et al \cite{shamout2021} that performed optimally over a 96-hour horizon compared to shorter 24h, 48h, and 72h time period, shortening the labeling time period could blur the distinction between positive and negative cases.  As a result, the model would need to discriminate more acutely between moderate and relatively severe OA on radiographs and MRI. 

The models developed in this work have the potential to hold a strong clinical utility once validated on clinical cohorts. The primary contribution of our work is the modified risk formulation with which the prognosis of a disease can be improved. Ideally, with a larger and longer longitudinal dataset, it is possible to use multiple scans to better prognose the patients (e.g., predict TKR within 9 years). With a better and more accurate prognosis earlier, patients can make lifestyle changes to slow down the progression of OA as much as they possibly can.

There are several limitations to our study. In both the OAI and MOST datasets, the knee radiographs were acquired in a systematic way, which might not be reflective of real-world imaging data. Further studies are needed to successfully translate the developed models into the real-world clinical setting where radiographs and MRI are acquired using different imaging systems and protocols.  Another limitation was that the datasets contained class imbalance due to the slow-progressing nature of OA. The much larger number of knees without TKR than those with TKR could limit model training and potentially affect model performances toward specific TKR case knees. Additionally, the decision for TKR surgery could be a biased endpoint due to the costs associated with the procedure, which may exclude individuals from certain socioeconomic groups. A more comprehensive endpoint could be a composite outcome measure combining TKR, pain scores, and radiographic scores. However, this approach also presents challenges. Subjectivity and varying pain thresholds among individuals can lead to noisy labels, making the task of training and evaluating models significantly more difficult. To mitigate these issues, methods for normalizing pain scores with respect to a baseline, possibly based on clinical factors, need to be developed. Additionally, while radiographic scores show a moderate correlation with TKR, this relationship requires further study to better understand its implications for predicting outcomes.

In conclusion, our study showed that the DL models trained with the proposed risk formulation had a higher performance in predicting TKR need over shorter time-horizons (1, 2, 4 years) than baseline DL models trained using conventional loss functions. The proposed model training method could benefit the research community and be applied to risk assessment models for other progressive diseases. Models trained for the prediction of progression-based outcomes both in and outside the domain of OA could incorporate this formulation to improve their performance and generalizability.






\section{ACKNOWLEDGMENTS}

This study was supported by National Institutes of Health (R01 AR074453).
The OAI is a public-private partnership comprised of five contracts (N01-AR-2-2258; N01-AR-2-2259; N01-AR-2-2260; N01-AR-2-2261; N01-AR-2-2262) funded by the National Institutes of Health, a branch of the Department of Health and Human Services, and conducted by the OAI Study Investigators. Private funding partners include Merck Research Laboratories; Novartis Pharmaceuticals Corporation, GlaxoSmithKline; and Pfizer, Inc. Private sector funding for the OAI is managed by the Foundation for the National Institutes of Health. This manuscript was prepared using an OAI public use data set and does not necessarily reflect the opinions or views of the OAI investigators, the NIH, or the private funding partners. MOST is a nationwide research study, sponsored by the National Institutes of Health / National Institute on Aging (part of the Department of Health and Human Services), that will help us better understand how to prevent and treat knee osteoarthritis, one of the most common causes of disability in adults. MOST is a collaborative effort developed by investigators at four core sites and funded by NIH/NIA grants: U01 AG18820 (David Felson, Boston University); U01 AG18947 (Cora E. Lewis, University of Alabama at Birmingham); U01 AG18832 (James Torner, University of Iowa); U01 AG19069 (Michael Nevitt, University of California, San Francisco).

\appendix
\label{appendix}
\section{ConReg}
In this setting for two scans  $x_1, x_2$ and model  $f$, the predicted representations at the penultimate layer
$(h_1, h_2)$ are obtained and used to compute the contrastive loss.

$$L_{cont} = y_{siam} * \|h_2 - h_1\|^2 + (1 - y_{siam}) * (\max(0, m - \|h_2 - h_1\|))^2$$

Where  $y_{siam}$ is the Siamese label with $y_{siam} = 1$ if $y_1 = y_2$ and $y_{siam} = 0$ if $y_1 \neq y_2$. So effectively the representations of the two scans are pushed together if their TKR labels are the same and they are pushed apart if their TKR labels are different. The total loss applied here is,

$$L_{total} = BCE(y_1, \hat{y}_1) + BCE(y_2, \hat{y}_2) + \gamma * L_{cont}$$

\section{RiskReg}

The loss regularization approach (RiskReg) to the application of risk constraint is formalized below. For
a knee, we have a scan from time step 1 ($x_1$), a scan from time step 2 ($x_2$) and a CNN model ($f$). The risk
of TKR event $E$ (within $t = 1,2,4$ years) for these two scans is given by,

$$p(E|x_1) = \hat{y}_1 = \sigma(f(x_1))$$

$$p(E|x_2) = \hat{y}_2 = \sigma(f(x_2))$$

where $\sigma$ is the logistic sigmoid function, $y_1, y_2 \in \{0,1\}$ indicating the true TKR status for
scans at time-step 1 and time-step 2 respectively, $t \in \{1,2,4\}$ in this study, $f(x_1), f(x_2) \in \mathbb{R}$,
$\sigma(f(x_1)), \sigma(f(x_2)) \in (0,1)$, and $\hat{y}_t^1, \hat{y}_t^2$ are the predicted TKR risks.

As stated earlier, our assumption is $p(E|x_1) \leq p(E|x_2)$, which translates to

$$\sigma(f(x_1)) \leq \sigma(f(x_2))$$

This constraint is enforced by adding a margin ranking loss term$^{14}$.

$$L_{reg} = max (0,log (\sigma(f(x_1))) -log (\sigma(f(x_2))) + m).$$

enforces the constraint $log (\sigma(f(x_1))) -log (\sigma(f(x_2))) \geq m$, where $m$ is the margin hyperparameter that controls how far apart the log sigmoid outputs should be (as $m$ is increased $\sigma(f(x_2))$ would be pushed farther away from $\sigma(f(x_1))$). The natural log (a strictly increasing function) is applied to the sigmoid output to achieve better numerical performance and avoid diminishing gradient problems associated with the sigmoid function during training. So, the total loss applied for a knee with two scans is,

$$L_{total} = BCE(y_1, \hat{y}_1) + BCE(y_2, \hat{y}_2) + \gamma * L_{reg}$$

Where $BCE$ is the standard binary cross entropy loss. For a knee with only one scan, the $L_{total} = BCE(y_1, \hat{y}_1)$ is applied during training.



\begin{thebibliography}{00}


\bibitem{cross2014}
Cross, Marita, et al.,
\textit{The global burden of hip and knee osteoarthritis: estimates from the global burden of disease 2010 study},
Annals of the rheumatic diseases 73.7,
2014.

\bibitem{tolpadi2020}
Tolpadi, Aniket A., et al.,
\textit{Deep learning predicts TKR from magnetic resonance images},
Scientific reports 10.1,
2020.

\bibitem{zhang2020}
Zhang, Bofei, et al.,
\textit{Attention-based cnn for kl grade classification: Data from the osteoarthritis initiative},
2020 IEEE 17th international symposium on biomedical imaging (ISBI),
IEEE,
2020.

\bibitem{leung2020}
Leung, Kevin, et al.,
\textit{Prediction of total knee replacement and diagnosis of osteoarthritis by using deep learning on knee radiographs: data from the osteoarthritis initiative},
Radiology 296.3,
2020.

\bibitem{dougados2009}
Dougados, Maxime, et al.,
\textit{OARSI/OMERACT criteria of being considered a candidate for total joint replacement in knee/hip osteoarthritis as an endpoint in clinical trials evaluating potential disease modifying osteoarthritic drugs},
The Journal of Rheumatology 36.9,
2009.

\bibitem{huang2017}
Huang, Gao, et al.,
\textit{Densely connected convolutional networks},
Proceedings of the IEEE conference on computer vision and pattern recognition,
2017.

\bibitem{tiulpin2018}
Tiulpin, Aleksei, et al.,
\textit{Automatic knee osteoarthritis diagnosis from plain radiographs: a deep learning-based approach},
Scientific reports 8.1,
2018.

\bibitem{tiulpin2020}
Tiulpin, Aleksei, and Simo Saarakkala,
\textit{Automatic grading of individual knee osteoarthritis features in plain radiographs using deep convolutional neural networks},
Diagnostics 10.11,
2020.

\bibitem{tiulpin2019a}
Tiulpin, Aleksei, et al.,
\textit{Multimodal machine learning-based knee osteoarthritis progression prediction from plain radiographs and clinical data},
Scientific reports 9.1,
2019.

\bibitem{tiulpin2019b}
Tiulpin, A., et al.,
\textit{Deep learning predicts knee osteoarthritis progression from plain radiographs},
Osteoarthritis and Cartilage 27,
2019.

\bibitem{pedoia2019}
Pedoia, Valentina, et al.,
\textit{Diagnosing osteoarthritis from T2 maps using deep learning: an analysis of the entire Osteoarthritis Initiative baseline cohort},
Osteoarthritis and cartilage 27.7,
2019.

\bibitem{rajamohan2023}
Rajamohan, Haresh Rengaraj, et al.,
\textit{Prediction of total knee replacement using deep learning analysis of knee MRI},
Scientific reports 13.1,
2023.

\bibitem{guan2022}
Guan, Bochen, et al.,
\textit{Deep learning approach to predict pain progression in knee osteoarthritis},
Skeletal radiology,
2022.

\bibitem{ningrum2021}
Ningrum, Dina Nur Anggraini, et al.,
\textit{A deep learning model to predict knee osteoarthritis based on nonimage longitudinal medical record},
Journal of Multidisciplinary Healthcare,
2021.

\bibitem{kim2020}
Kim, Dong Hyun, et al.,
\textit{Can additional patient information improve the diagnostic performance of deep learning for the interpretation of knee osteoarthritis severity},
Journal of Clinical Medicine 9.10,
2020.

\bibitem{wang2021}
Wang, Yifan, et al.,
\textit{An automatic knee osteoarthritis diagnosis method based on deep learning: data from the osteoarthritis initiative},
Journal of Healthcare Engineering 2021.1,
2021.

\bibitem{christodoulou2019}
Christodoulou, Eirini, et al.,
\textit{Exploring deep learning capabilities in knee osteoarthritis case study for classification},
2019 10th international conference on information, intelligence, systems and applications (IISA),
IEEE,
2019.

\bibitem{schiratti2021}
Schiratti, Jean-Baptiste, et al.,
\textit{A deep learning method for predicting knee osteoarthritis radiographic progression from MRI},
Arthritis Research \& Therapy 23,
2021.

\bibitem{abdullah2022}
Abdullah, S. Sheik, and M. Pallikonda Rajasekaran,
\textit{Automatic detection and classification of knee osteoarthritis using deep learning approach},
La radiologia medica 127.4,
2022.

\bibitem{ahmed2022a}
Ahmed, Sozan Mohammed, and Ramadhan J. Mstafa,
\textit{Identifying severity grading of knee osteoarthritis from x-ray images using an efficient mixture of deep learning and machine learning models},
Diagnostics 12.12,
2022.

\bibitem{swiecicki2021}
Swiecicki, Albert, et al.,
\textit{Deep learning-based algorithm for assessment of knee osteoarthritis severity in radiographs matches performance of radiologists},
Computers in biology and medicine 133,
2021.

\bibitem{song2023}
Song, Jiangling, and Rui Zhang,
\textit{A novel computer-assisted diagnosis method of knee osteoarthritis based on multivariate information and deep learning model},
Digital Signal Processing 133,
2023.

\bibitem{olsson2021}
Olsson, Simon, et al.,
\textit{Automating classification of osteoarthritis according to Kellgren-Lawrence in the knee using deep learning in an unfiltered adult population},
BMC musculoskeletal disorders 22,
2021.

\bibitem{gan2021}
Gan, Hong-Seng, et al.,
\textit{From classical to deep learning: review on cartilage and bone segmentation techniques in knee osteoarthritis research},
Artificial Intelligence Review 54.4,
2021.

\bibitem{hu2023}
Hu, Jiaping, et al.,
\textit{DeepKOA: a deep-learning model for predicting progression in knee osteoarthritis using multimodal magnetic resonance images from the osteoarthritis initiative},
Quantitative Imaging in Medicine and Surgery 13.8,
2023.

\bibitem{kokkotis2020}
Kokkotis, Christos, et al.,
\textit{Machine learning in knee osteoarthritis: A review},
Osteoarthritis and Cartilage Open 2.3,
2020.

\bibitem{bayramoglu2021}
Bayramoglu, Neslihan, Miika T. Nieminen, and Simo Saarakkala,
\textit{Automated detection of patellofemoral osteoarthritis from knee lateral view radiographs using deep learning: data from the Multicenter Osteoarthritis Study (MOST)},
Osteoarthritis and Cartilage 29.10,
2021.

\bibitem{rani2024}
Rani, Suman, et al.,
\textit{Deep learning to combat knee osteoarthritis and severity assessment by using CNN-based classification},
BMC Musculoskeletal Disorders 25.1,
2024.

\bibitem{mahum2023}
Mahum, Rabbia, et al.,
\textit{A robust framework for severity detection of knee osteoarthritis using an efficient deep learning model},
International Journal of Pattern Recognition and Artificial Intelligence 37.07,
2023.

\bibitem{lim2019}
Lim, Jihye, Jungyoon Kim, and Songhee Cheon,
\textit{A deep neural network-based method for early detection of osteoarthritis using statistical data},
International journal of environmental research and public health 16.7,
2019.

\bibitem{yeoh2021}
Yeoh, Pauline Shan Qing, et al.,
\textit{Emergence of deep learning in knee osteoarthritis diagnosis},
Computational intelligence and neuroscience 2021.1,
2021.

\bibitem{kijowski2023}
Kijowski, Richard, Jan Fritz, and Cem M. Deniz,
\textit{Deep learning applications in osteoarthritis imaging},
Skeletal radiology 52.11,
2023.

\bibitem{ahmed2022b}
Ahmed, Sozan Mohammed, and Ramadhan J. Mstafa,
\textit{A comprehensive survey on bone segmentation techniques in knee osteoarthritis research: From conventional methods to deep learning},
Diagnostics 12.3,
2022.

\bibitem{zhao2025}
Zhao, Haoming, et al.,
\textit{The value of deep learning-based X-ray techniques in detecting and classifying KL grades of knee osteoarthritis: a systematic review and meta-analysis},
European Radiology 35.1,
2025.

\bibitem{li2020}
Li, Matthew D., et al.,
\textit{Siamese neural networks for continuous disease severity evaluation and change detection in medical imaging},
NPJ digital medicine 3.1,
2020.

\bibitem{karsdal2016}
Karsdal, M. A., et al.,
\textit{Disease-modifying treatments for osteoarthritis (DMOADs) of the knee and hip: lessons learned from failures and opportunities for the future},
Osteoarthritis and cartilage 24.12,
2016.

\bibitem{nevitt2006}
Nevitt, Michael, David Felson, and Gayle Lester,
\textit{The osteoarthritis initiative},
Protocol for the cohort study 1,
2006.

\bibitem{segal2013}
Segal, Neil A., et al.,
\textit{The Multicenter Osteoarthritis Study (MOST): opportunities for rehabilitation research},
PM \& R: the journal of injury, function, and rehabilitation 5.8,
2013.

\bibitem{kellgren1957}
Kellgren, Jonas H., and JS1006995 Lawrence,
\textit{Radiological assessment of osteo-arthrosis},
Annals of the rheumatic diseases 16.4,
1957.

\bibitem{he2016}
He, Kaiming, et al.,
\textit{Deep residual learning for image recognition},
Proceedings of the IEEE conference on computer vision and pattern recognition,
2016.

\bibitem{he2015}
He, Kaiming, et al.,
\textit{Delving deep into rectifiers: Surpassing human-level performance on imagenet classification},
Proceedings of the IEEE international conference on computer vision,
2015.

\bibitem{kingma2014}
Kingma, Diederik P., and Jimmy Ba,
\textit{Adam: A method for stochastic optimization},
arXiv preprint arXiv:1412.6980,
2014.

\bibitem{delong1988}
DeLong, E. R., DeLong, D. M. \& Clarke-Pearson, D. L.,
\textit{Comparing the areas under two or more correlated receiver operating characteristic curves: A nonparametric approach},
Biometrics 44(3),
1988.
\bibitem{shamout2021}
Shamout, Farah E., et al.,
\textit{An artificial intelligence system for predicting the deterioration of COVID-19 patients in the emergency department},
NPJ digital medicine 4.1,
2021.

\bibitem{weston1999}
Weston, Jason, and Chris Watkins,
\textit{Support vector machines for multi-class pattern recognition},
Esann, Vol. 99,
1999.

\bibitem{kim2023}
Kim, Heejong, and Mert R. Sabuncu,
\textit{Learning to compare longitudinal images},
arXiv preprint arXiv:2304.02531,
2023.

\end{thebibliography}



\end{document}